\documentclass[a4paper,12pt]{article}

\usepackage{amsmath}
\usepackage{amssymb}
\usepackage{latexsym}
\usepackage{cite}
\usepackage[dvips]{graphicx}
\usepackage{bbold}
\usepackage{epsfig}
\usepackage{amsfonts}

\newcommand{\beqa}{\begin{eqnarray}}
\newcommand{\eeqa}{\end{eqnarray}}
\newcommand{\bea}{\begin{eqnarray}}
\newcommand{\eea}{\end{eqnarray}}

\newcommand{\be}{\begin{equation}}
\newcommand{\ee}{\end{equation}}
\newcommand{\half}{\frac{1}{2}}
\newcommand{\tr}{\,\textup{tr}}

\newcommand{\mcA}{{\mathcal A}}
\newcommand{\mcB}{{\mathcal B}}
\newcommand{\mcC}{{\mathcal C}}

\newcommand{\mcF}{{\mathcal F}}

\newcommand{\mcH}{{\mathcal H}}

\newcommand{\mcM}{{\mathcal M}}
\newcommand{\mcN}{{\mathcal N}}

\newcommand{\mcR}{{\mathcal R}}

\newcommand{\mcK}{{\mathcal K}}
\renewcommand{\Re}{\textup{Re}}
\renewcommand{\Im}{\textup{Im}}

\newcommand{\bj}{\bar j}
\newcommand{\bi}{\bar i}
\newcommand{\bz}{\bar z}
\newcommand{\bw}{\bar w}

\numberwithin{equation}{section}

\begin{document}
\title{\bf Hermitian Yang-Mills instantons\\ on Calabi-Yau cones}\vspace{6pt}
\author{\normalsize{Filipe Paccetti Correia\footnote{paccetti@fc.up.pt}} \vspace{6pt}\\
{{\it \small{Centro de F\'\i sica do Porto}}},
\\ {{\it \small{Faculdade de Ci\^encias da Universidade do Porto}}},
\\ {{\it \small{Rua do Campo Alegre, 687, 4169-007 Porto, Portugal}}}}

\date{\normalsize{October 6, 2009}}

\maketitle

\abstract{We study and construct non-abelian hermitian Yang-Mills (HYM) instantons on Calabi-Yau cones. By means of a particular isometry preserving ansatz, the HYM equations are reduced to a novel Higgs-Yang-Mills flow on the Einstein-K\"ahler base. For any 2$d_{\mathbb C}$-dimensional Calabi-Yau cone, we find explicit solutions of the flow equations that correspond to non-trivial SU($d_{\mathbb C}$) HYM instantons. These can be regarded as deformations of the spin connection of the Calabi-Yau cone.}

\section{Introduction}


The lack of explicit solutions can sometimes hinder the progress of a given field of theoretical physics. This is to some extent the case with Calabi-Yau (CY) compactifications of the heterotic string, which are of interest for obtaining Grand Unified Theories. In this context, one is required to specify both a (compact) CY manifold and a non-trivial non-abelian HYM instanton on that manifold \cite{Candelas:1985en}. It is a well known problem that neither the Ricci-flat metrics on the CY manifolds nor the non-abelian HYM instantons over them are explicitly known. This forces us to study only those features of the effective low-energy theory that are determined by the topological properties of the CY manifold and the HYM instanton. One of the reasons for this state of affairs is that \emph{compact} CY manifolds do not possess continuous isometries. 

In this paper we approach this problem by studying HYM instantons over \emph{non-compact} Ricci-flat K\"ahler cones, dubbed Calabi-Yau cones in the following. This is, of course, a drastic simplification of the problem, but it might give us some intuition on what happens in (certain limits of) the compact case. One of the main values of this approach lies in the fact that infinitely many explicit metrics on CY cones \cite{Gauntlett:2004yd,Gauntlett:2004hh,Cvetic:2005ft,Cvetic:2005vk,Lu:2005sn} have been constructed in recent years. In this paper we develop a recipe that will allow us to construct a 1-parameter family of SU($d_{\mathbb C}$) HYM instantons on any CY cone $d_{\mathbb C}$-fold, given the spin-connection of the cone. Each of these solutions is a deformation of the spin-connection, to which it converges at the apex of the cone.

In contrast to the compact case, Ricci-flat metrics on CY cones always possess at least one continuous isometry - generated by the so-called Reeb vector field. This naturally suggests an ansatz for the HYM connection that preserves this symmetry in the presence of the instanton background too. As we shall show in this paper, the resulting hermitian YM equations can then be interpreted as describing a new Higgs-Yang-Mills \emph{flow}, with the radial coordinate of the CY geometry assuming the r\^ole of the flow parameter. This flow takes place on the compact Einstein-K\"ahler (EK) base of the CY cone, and - for bounded and non-trivial instantons - interpolates between two \emph{distinct} YM instantons, both of them extrema of the Yang-Mills functional on the EK space. There is also an adjoint Higgs-field whose value \emph{must} change under the flow. The explicit SU($d_{\mathbb C}$) instantons over CY cone ${d_{\mathbb C}}$-folds presented in this paper are \emph{particular} solutions of these flow equations, constructed using the spin-connections of the EK bases. Instanton solutions for other symmetry groups can also be constructed explicitly and will be presented elsewhere \cite{paccetti2009}.

From the viewpoint of heterotic compactifications, the present work should be seen as a first step towards a more ambitious goal, namely the construction of HYM instantons over resolutions of CY cones. It would be an interesting challenge to determine the moduli space parametrising the \emph{combined} deformations of the resolved CY cone and of the HYM instanton. We would then be in the position of obtaining and studying the K\"ahler potentials for a large class of 4d heterotic theories, thus going beyond the single known example \cite{Gray:2003vw} of a K\"ahler potential that includes the instanton moduli.\\

The paper is organised as follows. We first review, in section \ref{sec:CY_cones}, several useful facts about the geometry of K\"ahler and Calabi-Yau cones. This is followed by a discussion of the hermitean Yang-Mills equations on Calabi-Yau cones in section \ref{sec:HYM_on_CYcones}. There, we present an ansatz that trades the HYM equations on CY cones for Higgs-Yang-Mills flow equations on Einstein-K\"ahler spaces. In section \ref{sec:SU(n)instantons}, we find and discuss a class of solutions to the aforementioned flow equations. Concrete examples of such solutions are presented for three different CY cones in section \ref{sec:examples}. We close with a few remarks on open problems and directions for future research.

\section{Calabi-Yau cones}\label{sec:CY_cones}

We are interested on HYM instantons on Calabi-Yau cones, that is Ricci flat K\"ahler cones. We begin by reviewing some generalities on K\"ahler spaces and setting up the notation, before specializing to the cones.

\subsection{K\"ahler spaces}

A K\"ahler metric can always be locally written as
\be
             ds^2=2\mcK_{i\bj}~dz^i\otimes d\bz^{\bj} \ ,
\ee
where $\mcK_{i\bj}=\partial_i\partial_{\bj}\mcK(z,\bz)$ and $\mcK(z,\bz)$ is the K\"ahler potential. Clearly, this implies that the K\"ahler 2-form,
\be
             J=-i\mcK_{i\bj}~dz^i\wedge d\bz^{\bj} \ ,
\ee
is closed. Let us introduce an orthonormal frame
\be
             e_i=g^{\dagger}_{ij}dz^j \ , 
\ee
defined in terms of a squared matrix $g$ of rank $d_{\mathbb C}$ satisfying
\be
             (gg^{\dagger})_{\bj i}=\mcK_{i\bj} \ ,
\ee
such that now
\be
             ds^2=2 e_i\otimes {\bar e}_{\bi} \ , \quad J=-i e_i\wedge {\bar e}_{\bi} \ . 
\ee
The spin-connection $\Omega$, determined by
\be
             de_i+\Omega_i^{~k}\wedge e_k=0 \ ,
\ee
can be shown to read
\be\label{eq:def_spin_connect}
             \Omega=g^{-1}\partial g+g^{\dagger}\bar{\partial}g^{\dagger-1} \ ,
\ee
\emph{for a K\"ahler space}. This means that $\Omega$ is the connection of a holomorphic vector bundle. It is useful to split the spin-connection into its holomorphic and anti-holomorphic parts
\be
             \Omega=\mcA+\bar{\mcA} \ ,
\ee
with $\mcA=g^{-1}\partial g$ and $\bar{\mcA}=-\mcA^{\dagger}$. The holomorphy of the tangent bundle is then equivalent to the fact that
\be
             \mcF_{2,0}(\mcA)=\partial\mcA+\mcA\wedge\mcA=0 \ .
\ee
Then, the only non-vanishing piece of the tangent bundle's curvature 2-form is the $(1,1)$-component that reads
\be
             \mcR(\Omega)=\mcF_{1,1}(\mcA,\bar{\mcA})=\bar{\partial}\mcA+\partial{\bar\mcA}+\mcA\wedge\bar{\mcA}+\bar{\mcA}\wedge\mcA \ .
\ee
For future use we also note that $\mcR_i^{~j}\wedge e_j=0$.

The curvature 2-forms of Einstein-K\"ahler spaces have the particularity of satisfying the so-called hermitian Yang-Mills equation (HYM),
\be
             \mcK^{i\bj}\mcR_{i\bj}^{ab}=\frac{R}{2d_{\mathbb C}}~\delta^{ab} \ ,
\ee
with $R=-2\mcK^{i\bj}\partial_i\partial_{\bj}\ln\det (\mcK_{a\bar{b}})$ being the (constant) scalar curvature. This can also be neatly rephrased as 
\be
             J^{d_{\mathbb C}-1}\wedge\mcR=i{\mathbb 1}\frac{R}{2d^2_{\mathbb C}}J^{d_{\mathbb C}} \ ,
\ee
which, in the particular case of a Calabi-Yau $d_{\mathbb C}$-fold reads
\be
             J^{d_{\mathbb C}-1}\wedge\mcR=0 \ .
\ee

\subsection{K\"ahler cones}
A rather well studied class of non-compact K\"ahler spaces is that of Ricci-flat K\"ahler cones, with an infinitely large number of explicit metrics being known by now  \cite{Gauntlett:2004yd,Gauntlett:2004hh,Cvetic:2005ft,Cvetic:2005vk,Lu:2005sn}. By definition, any $2d_{\mathbb C}=2(n+1)$-dimensional K\"ahler cone is a cone over a $(2n+1)$-dimensional Sasaki space, which in turn is a line bundle over a $2n$-dimensional K\"ahler base. In the following, we introduce local complex coordinates $w^i$ ($i=1,\dots,n$) for the latter
and denote its K\"ahler potential by $\mcK(w,\bw)$. The K\"ahler potential for the cone can then be written as 
\be
              \mcK_C=|z|^2e^{2\mcK(w,\bw)}\equiv\half\rho^2 \ ,
\ee
where $z\in{\mathbb C}$ vanishes at the apex of the cone. With $z=\rho e^{-\mcK}e^{i\phi}/\sqrt{2}$, we find that the metric of a K\"ahler cone,
\be
              ds^2_C=d\rho^2+\rho^2ds^2(Y) \ , \quad \rho>0 \ ,
\ee
is determined by the metric of a Sasaki space $Y$
\be
              ds^2(Y)=\eta\otimes\eta+2\mcK_{i\bj}dw^id\bw^{\bj} \ ,
\ee
where the 1-form
\be
              \eta=d\phi-i(\partial-\bar{\partial})\mcK \ ,
\ee
is the dual of the \emph{Reeb} Killing vector field. Clearly, the real coordinate $\rho$ has the interpretation of a radial coordinate measuring the distance to the apex of the cone. 

The 1-form $\eta$ determines the K\"ahler 2-form of the $2n$ dimensional base,
\be
              J=-\half d\eta \ ,
\ee
as well as the K\"ahler form of the cone,
\be
              J_C=-\half d(\rho^2\eta) \ .
\ee
Likewise, the curvature of the cone is fully determined by the K\"ahler base geometry. For instance, the Ricci 2-form reads
\be\label{eq:Ricci_cone}
              \mcR ic_C=2(n+1)J+\mcR ic \ ,
\ee
while the scalar curvature
\be
              R_C=\frac{R-4n(n+1)}{\rho^2} \ .
\ee
Hence, imposing the Calabi-Yau condition on the cone implies that the base space is Einstein-K\"ahler with the scalar curvature being set by the dimensionality of the cone as 
\be
              R=4n(n+1) \ .
\ee
Equivalently, one can state that the K\"ahler base has $U(n)$ holonomy and its curvature is HYM with 
\be\label{eq:HYM_for_basetangent}
             J^{n-1}\wedge\mcR=i{\mathbb 1}\frac{2(n+1)}{n}J^n \ ,
\ee
(or $\mcK^{i\bj}\mcR_{i\bj}^{ab}=2(n+1)\delta^{ab}$).

The HYM instantons that we will present in section \ref{sec:SU(n)instantons} can be regarded as deformations of the CY cone's spin-connection $\Omega_C$. As such, it is useful to have an expression for $\Omega_C$ in terms of the spin-connection on the Einstein-K\"ahler base. Recall that $\Omega_C$ is determined by a rank $(n+1)$ square matrix $g$ (cf. Eq.\eqref{eq:def_spin_connect}), which in the case of a cone can be written as
\be
             g=e^{\mcK}\left(\begin{array}{cc}
              -1 &\quad 0 \\
              -2\bz\mcK_{\bar{a}} & \sqrt{2}\bz h_{\bar{a}b}
              \end{array} \right) \ ,
\ee
where $h$ is a rank $n$ matrix determining the spin-connection of the K\"ahler base. (In particular we have $(hh^{\dagger})^{\bar{b} a}=\mcK_{a\bar{b}}$.) The holomorphic part of $\Omega_C$ is then straightforward to compute
\be\label{eq:spin_connec_cone}
             \mcA_C=g^{-1}\partial g=\left(\begin{array}{cc}
              \partial\mcK &\quad 0 \\
              -\sqrt{2}e_b & \mcC_b^{~a}+ \delta_b^{~a}\partial\mcK
               \end{array} \right) \ ,
\ee
where, for convenience, we introduced the orthonormal frame $e_i=h^{\dagger}_{ia}dw^a$, and $\mcC=h^{-1}\partial h$ is the (holomorphic part of the) U($n$) spin-connection on the K\"ahler base space. We should note that $\mcA_C$ is determined only up to gauge transformations which, by definition, are unitary transformations acting on $g$ from the left, leaving $gg^{\dagger}$ invariant. 

Imposing that the cone is Ricci-flat implies $\mcA_C$ to be an SU($n+1$) connection. In fact, for $\mcC$ this implies that
\be   
              \tr(\mcC)=-(n+1)\partial\mcK \ ,
\ee
and thus
\be
              \mcR ic\equiv i\tr(\mcR)=i2(n+1)\partial\bar{\partial}\mcK=-2(n+1)J \ .
\ee
Then, upon the use of Eq.\eqref{eq:Ricci_cone} this implies the CY condition, $\mcR ic_C=0$, to be satisfied on the CY cone as it should.

In the gauge we are using here, the curvature of the tangent bundle of a cone is straightforward to compute and gives
\be\label{eq:Tang_bundle_curvature}
              \mcR_C=\left(\begin{array}{cc}
              0 &\quad 0 \\
              0 & \mcR^{b\bar{a}}-2(\delta^{i\bj}\delta^{b\bar{a}}+\delta^{\bar{a}\bj}\delta^{bi})e_i\wedge\bar{e}_{\bj}
              \end{array} \right) \ .
\ee
Below, we will discuss HYM instantons over CY cones. In particular, we will see that if \eqref{eq:Tang_bundle_curvature} is to satisfy the HYM equation (i.e. $J_C^n\wedge\mcR_C=0$) one should then have 
\be
             J^{n-1}\wedge(\mcR^{b\bar{a}}-2(\delta^{i\bj}\delta^{b\bar{a}}+\delta^{\bar{a}\bj}\delta^{bi})e_i\wedge\bar{e}_{\bj})=0 \ .
\ee
This is, indeed, equivalent to Eq.\eqref{eq:HYM_for_basetangent}.

\section{The HYM equation on Calabi-Yau cones}\label{sec:HYM_on_CYcones}

Compactifications of the heterotic string with 4d $\mcN=1$ supersymmetry at low energies require, to lowest order in $\alpha'$, selecting both a Calabi-Yau 3-fold and a hermitean Yang-Mills instanton over the Calabi-Yau 3-fold. In this section, after presenting the HYM equations for CY cones, we will use a particular ansatz to suggestively rewrite the HYM equations as a (novel) Higgs-Yang-Mills flow. We will also derive a few general features of any instanton that is a solution of these equations.

\subsection{Holomorphy and DUY conditions}
Given a Calabi-Yau manifold of complex dimension $d_{\mathbb{C}}=n+1$, an hermitian Yang-Mills instanton with curvature $\mcF$ and $c_1(\mcF)=0$ must satisfy
\be\label{eq:holomorphy}
                  \qquad \mcF_{(2,0)}=0  \quad\textup{(holomorphy)}
\ee
\be\label{eq:DUY}
                J_{\mcC}^n\wedge\mcF_{(1,1)}=0  \quad\textup{(DUY equation)}
\ee
The holomorphy condition and the Donaldson-Uhlenbeck-Yau equation are sometimes also referred to as F-term and D-term conditions, respectively. In this paper, we want to find and study solutions of these equations over generic Calabi-Yau cones. We use the results of the previous section to write $J_{\mcC}^n$ as
\be
                 J_{\mcC}^n=\rho^{2n}J^{n}-n\rho^{2n-1}d\rho\wedge\eta\wedge J^{n-1} \ .
\ee
It then follows that any solution $\mcF$ of the DUY equation (Eq.\eqref{eq:DUY}) must be of the general form
\be\label{eq:master_cond}
                 \mcF=\mcH+F\left(\epsilon\wedge\bar{\epsilon}-\frac{1}{n}2\mcK_{i\bar j}dw^i\wedge d\bar{w}^{\bar j} \right) + \bar{\beta}\wedge\epsilon+\beta\wedge\bar{\epsilon} \ ,
\ee
where $\mcH,~\beta$, and $\bar{\beta}$ are respectively $(1,1)$, $(1,0)$ and $(0,1)$-forms tangent to the base of the cone, $F$ is a function, and we introduced 
\be
                 \epsilon=\frac{dz}{z}+2\mcK_i dw^i=\frac{d\rho}{\rho}+i\eta \ .
\ee
Moreover $\mcH$ is constrained to satisfy
\be
                 J^{n-1}\wedge \mcH=0 \ .
\ee
The latter equation can be interpreted as a DUY equation on the Einstein-K\"ahler n-fold (spanned by the coordinates $w^i,\bw^{\bj}$) that must be satisfied fiberwise at any value of $z$. This fact motivates us to attempt a construction of HYM instantons on CY cones using instantons on the Einstein-K\"ahler space as seeds.

\subsection{An Ansatz and a flow}

Here, we consider a simplifying but natural ansatz for the gauge instanton. Namely, we assume that the gauge instanton does not depend on the coordinate $\phi$ related with a natural isometry of the CY cone. To be more precise, we write the gauge connection as
\be
           \mcA=\mcA_{\epsilon}(t,w,\bar{w})\epsilon+\mcB_i(t,w,\bar{w})dw^i \ ,
\ee 
where $t=\ln\rho^2$. The holomorphy condition $\mcF_{2,0}(\mcA)=0$ then decomposes as
\be
             \mcF_{2,0}(\mcB)=0 \ ,
\ee
and
\be\label{eq:Fterm_2}
             \partial_t\mcB=D_{\mcB}(\Phi+iA_t) \ ,
\ee
where $\Phi=\Re(\mcA_{\epsilon})$, $A_t=\Im(\mcA_{\epsilon})$ and $D_{\mcB}=\partial+[\mcB,\cdot]$. We note that while $\Phi$ can be regarded as an adjoint Higgs field, $A_t$ transforms as a component of a \emph{real} gauge field under ($\phi$-independent) gauge transformations. This is relevant for then we can use an axial gauge choice to set $A_t=0$, thus simplifying Eq.\eqref{eq:Fterm_2}. On the other hand we find
\be\label{eq:F(t)}
            F=-2\partial_t\Phi \ ,
\ee
\be\label{eq:H(t)}
            \mcH=\mcF_{1,1}(\mcB)-\frac{2}{n}\left[\partial_t\Phi+n\Phi\right]2{\mcK}_{i\bar j}dw^i\wedge d\bar{w}^{\bar j} \ ,
\ee
and
\be\label{eq:beta(t)}
            \beta=-\partial_t\mcB-D_{\mcB}\Phi \ ,\quad \bar{\beta}=-\beta^{\dagger} \ .
\ee
To summarize, assuming a $\phi$-independent gauge field as well as the axial gauge, the equations for HYM SU(N) instantons on CY cones read
\be\label{eq:Fterm_3a}
             \mcF_{2,0}(\mcB)=0 \ ,
\ee
\be\label{eq:Fterm_3b}
             \partial_t\mcB=D_{\mcB}\Phi \ ,
\ee
and
\be\label{eq:Dterm_1}
             \partial_t\Phi=-n\Phi+\tfrac{1}{4}\mcK^{i\bar j}\mcF_{i\bar j} \ .
\ee
It is interesting to notice that these equations describe a Higgs-Yang-Mills flow, parametrised by $t$, on the Einstein-K\"ahler base manifold. Recall that in the Yang-Mills flow of \cite{atiyah_bott_1982,donaldson_1985}, instead of \eqref{eq:Dterm_1} one has
\be
            \Phi=-\mcK^{i\bar j}\mcF_{i\bar j} \ ,
\ee
leading to a flow equation for holomorphic connections
\be
             \partial_t\mcB=-D_{\mcB}(\mcK^{i\bar j}\mcF_{i\bar j}) \ .
\ee 
It is well known that this implies the Yang-Mills flow to be a gradient (or heat) flow for the Yang-Mills functional $\int \tr(\mcK^{i\bar j}\mcF_{i\bar j})^2$. It has the nice property that the gauge connection flows to a fix point that minimises the Yang-Mills functional. 

The flow we introduced above has nice properties of its own, as we now describe. We must require that our solutions give rise to \emph{finite} instanton numbers. This imposes that at $t=\pm\infty$, i.e. both at the apex of the cone and at asymptotic infinity, we must have $\partial_t\Phi=\partial_t\mcB=0$. In other words, the flow we are searching for must interpolate between two fix points of the above flow equations. We shall denote the fix point at $t=+\infty$ UV fix point, the one at $t=-\infty$ IR fix point. Now, the fix points satisfy the following set of equations
\be
             \mcF_{2,0}(\mcB)=0 \ ,
\ee
\be
             D_{\mcB}(\mcK^{i\bar j}\mcF_{i\bar j})=0 \ ,
\ee
and
\be\label{eq:fix_point_3}
            \Phi = \tfrac{1}{4n}\mcK^{i\bar j}\mcF_{i\bar j} \ .
\ee
Further insight on the flow can be obtained by looking at the functional
\be\label{eq:def_M(t)}
              M(t)=(\partial_t+n)\int_{EK_n}\tr(\Phi^2) \ ,
\ee
which can easily be shown to satisfy 
\be\label{eq:for_M(t)}
              \partial_t M(t)=\int_{EK_n}\tr\left(2(\partial_t\Phi)^2+|D_ {\mcB}\Phi|^2\right) \geq 0 \ .
\ee
Clearly, according to \eqref{eq:fix_point_3}, at the fix points $M(t)$ is proportional to the Yang-Mills functional. Moreover, Eqs.\eqref{eq:def_M(t)} and \eqref{eq:for_M(t)} imply that $M(t)$ is a \emph{growing, semi-positive} function of $t$. Semi-positiveness follows from noting that at the IR we have $M_{IR}=n\int_{EK_n}\tr(\Phi^2(-\infty))\geq 0$ and $M(t)$ is growing with $t$. Then, we also find that $\Phi_{UV}\equiv\Phi(t=+\infty)\neq 0$, otherwise $\partial_t\Phi=\partial_t\mcB=0$ everywhere and the instanton would be uninteresting in $(n+1)$-complex dimensions. 

The upshot of this discussion is that any interesting (\emph{i.e.} non-trivial and bounded) solution of the above set of equations must interpolate between one extremum of the Yang-Mills functional over the Einstein-K\"ahler space at asymptotic infinity (UV), and a second extremum with a smaller value of the same functional at the apex (IR). Recall that an hermitian Yang-Mills connection is an absolute minimum of the Yang-Mills functional, which vanishes there (for semi-simple Lie groups). As such, it is conceivable that at the apex the \emph{holomorphic} connection $\mcB$ approaches a HYM connection over the EK base manifold. In the SU($n+1$) case, there is an obvious such connection, namely the spin-connection of the Calabi-Yau cone. In the following section, for each CY cone, we will be able to construct smooth SU($n+1$) instantons that, indeed, are described by a flow from a \emph{non}-HYM, holomorphic solution of the Yang-Mills equations over the EK base at the UV to the spin-connection at the IR. Likewise, we will see that at the UV the adjoint "Higgs" $\Phi$ has a non-vanishing VEV that flows to zero at the IR.

\section{SU($n+1$) HYM instantons}\label{sec:SU(n)instantons}

We shall consider the simplest SU($n+1$) ansatz to solve the above flow equations on a CY cone $(n+1)$-fold. That is, we assume that 
\be
               \Phi=\varphi(t)\Sigma \ , \qquad  \Sigma=\left(\begin{array}{cc}
              1 &\quad 0 \\
              0 & -\frac{1}{n}\mathbb{1}_{n\times n}
              \end{array} \right) \ ,
\ee
which readily implies that $\partial_t\mcB=[\mcB,\Phi]$. By contrast, for $\mcB$ we still take the most general possible ansatz, with
\be
              \mcB=\left(\begin{array}{cc}
              \alpha &\quad E_{-} \\
              E_+ & \Gamma_{n\times n}
              \end{array} \right) \ .
\ee
Eq.\eqref{eq:Fterm_3b} then implies that $\partial_t \alpha=0=\partial_t\Gamma$, while
\be
            \partial_t E_{\pm}=\pm\frac{n+1}{n}\varphi(t) E_{\pm} \ ,
\ee
and thus
\be
            E_{\pm}(t,w,\bw)=e^{\pm\frac{n+1}{n}\int^t \varphi dt}E_{\pm}^0(w,\bw) \ .
\ee
According to the discussion at the end of previous section, for any non-trivial instanton we must have $\Phi_{UV}\neq 0$. It then follows that also $\varphi_{UV}\neq 0$ and either $E_-$ or $E_+$ must diverge at $t=+\infty$, unless one of these vanishes everywhere. To avoid a singular instanton at the $UV$, we will take $E_-^0=0$, denoting $E_+=E$ from now on. Notice that one could instead set $E_+^0=0$, but this would simply amount to a different gauge choice.  

Getting back to the condition that $\mcB$ should be holomorphic ($\mcF_{2,0}(\mcB)=0$), we learn that
\be
           \partial\alpha=0 \ , \qquad \partial\Gamma+\Gamma\wedge\Gamma=0 \ ,
\ee
and
\be
           \partial E^0+E^0\wedge\alpha+\Gamma\wedge E^0=0 \ .
\ee
We already know one solution to this set of equations, given by the components of the CY cone's spin-connection $\mcA_C$ (see Eq.\eqref{eq:spin_connec_cone}). We shall consider this possibility here, leaving a discussion on the general situation to the appendix. We thus set
\be\label{eq:solution_base}
           \alpha=\partial\mcK \ ,\quad E_{b}^0=-\sqrt{2}~e_b \ , \quad \Gamma_b^{~a}=\mcC_b^{~a}+\delta_b^{~a}\partial\mcK \ .
\ee
It is then not difficult to compute the curvature of $\mcB$ along the EK base,
\be
           \mcF(\mcB)=\mcR_C+2\left(\exp\left(2\dfrac{n+1}{n}\int\varphi dt\right)-1\right)\left(\begin{array}{cc}
              e_i\wedge\bar{e}_i &\quad 0 \\
              0 & -e_b\wedge\bar{e}_a
              \end{array} \right) \ ,
\ee
and to find that 
\be
           \mcK^{i\bj}\mcF_{i\bj}(\mcB)=2n\left[\exp\left(2\dfrac{n+1}{n}\int\varphi dt\right)-1\right]\Sigma \ .
\ee
Plugging this back in Eq.\eqref{eq:Dterm_1}, we finally obtain the flow equation for $\varphi(t)$,
\be\label{eq:flow_phi}
           \partial_t\varphi+n\varphi+\frac{n}{2}\left(1-\exp\left(2\dfrac{n+1}{n}\int\varphi dt\right)\right)=0 \ .
\ee
This equation determines the radial dependence of our HYM instanton, and as such it is important to understand its solutions. We can get some insight by recasting Eq.\eqref{eq:flow_phi} as
\be\label{eq:PDE_profile}
         X''+X'=-\frac{d}{dX}\left(X-Me^{X/M}\right) \ ,
\ee
where $X=2n\int\varphi dt$, primes denote differentiation with respect to $\tau=nt$, and $M=n^2/(n+1)$. Equation \eqref{eq:PDE_profile} describes the motion of a damped particle in the potential $V(X)=X-Me^{X/M}$. This potential has a maximum at $X=0$, and two runaway directions towards $X\to\pm\infty$. The solutions of interest, that is the solutions that lead to instantons of bounded curvature and finite instanton numbers, are those that start at the maximum at $\tau=-\infty$ ($X_{IR}=0$) and have $X(\tau)<0$ for finite $\tau$. The case of 4d CY cones ($n=1$), i.e. when the base is $\mathbb{CP}^1$ endowed with an Einstein-K\"ahler metric, is simple to solve, for in this case \eqref{eq:PDE_profile} reduces to the first order equation
\be 
            X'=-1+e^X \ .
\ee
The solution depends on a single parameter $t_0$ and reads
\be
            X=-\ln(1+e^{t-t_0}) \ , \qquad \varphi=-\tfrac{1}{2}\frac{1}{1+e^{-(t-t_0)}} \ .
\ee
The parameter $t_0$ can be thought of as the instanton's size modulus.

The asymptotic behaviour of $\varphi=X'/2$ is relevant for the computation of the instanton numbers and its form for general $n$ can easily be found
\be
          \varphi=\left\{\begin{array}{cc}
              -|c|e^t & ,~ t\to -\infty \\
              -\half~ & ,~ t\to +\infty
              \end{array} \right. \ .
\ee
In more detail, the large $t$ asymptotic behaviour for $n\geq 2$ is
\be
            X=-n(t-t_0)-\frac{M^2}{M-1}e^{-\frac{n+1}{n}(t-t_0)}+\cdots
\ee
\be
            \varphi=-\half+\half\frac{M}{M-1}e^{-\frac{n+1}{n}(t-t_0)}+\cdots
\ee
As it is manifest here, apart from the dimension of the EK base $n$, the instanton solutions presented above depend only on one parameter, the instanton's size $t_0$. In the large instanton limit ($t_0=+\infty$) we recover the spin connection of the CY cone, while in the small instanton limit ($t_0=-\infty$) there is a transition from an SU($n+1$) down to an SU($n$)$\times$U(1) instanton. At this so-called \emph{small instanton transition}, the second Chern character at $z=0$ undergoes a discontinuous change, decreasing by
\be\label{eq:inst_disc}
                 \frac{1}{2\pi^2}\frac{n+1}{n}J\wedge J \ ,
\ee
and the third Chern class jumps to zero. These are the mathematical facts, but there is a physical side to this story too. Recall that in heterotic string compactifications the second Chern character of the instanton is as a source of 3-form flux that must be balanced against the second Chern character of the background CY geometry in order that the configuration preserves a minimum of supersymmetry and is stable. One might then worry that the theory stops making sense in the small instanton limit, unless some physical process introduces a new source of 3-form flux to compensate the removal of (part of) the gauge instanton at $z=0$. In 6d compactifications, this physical process turns out to be the nucleation of a 5-brane wrapped on the holomorphic cycle dual to the 4-form \eqref{eq:inst_disc}. This has been elucidated in \cite{Witten:1995gx} and discussed with specific examples e.g. in \cite{Kachru:1997rs,Ovrut:2000qi}.


It is instructive to go back and describe our SU($n+1$) instantons from the point of view of the Higgs-Yang-Mills flow on the Einstein-K\"ahler n-fold. The flow starts at the UV fix point with an SU($n$)$\times$U(1), \emph{non-hermitian}, instanton over the EK$_n$ base, and an \emph{adjoint Higgs} $\Phi$ with a non-vanishing VEV. This VEV is covariantly flat. Once we are an infinitesimal distance away from the UV fix point, the adjoint Higgs does break the gauge symmetry, and this induces a flow of the gauge field $\mcB$ that now becomes a fully-fledged SU($n+1$) holomorphic connection on EK$_n$. It is not difficult to check that under the flow the Yang-Mills functional is monotonically decreasing. As such, at the IR one attains the absolute minimum of the Yang-Mills functional and $\Phi=0$. At this (fix) point the gauge field $\mcB$ becomes the connection of an HYM instanton over the EK$_n$ base space, but now with SU($n$) gauge symmetry. Notice that the gauge bundle over the EK$_n$ base undergoes two relevant changes. It starts at the UV as the tensor product of a rank-$n$ bundle with a line bundle. Then it is deformed and flows as a rank-$(n+1)$ gauge bundle until its rank is reduced by one unit at the IR. There are no other changes in the topological data in the sense that 
\be
                \tr\mcF(\mcB)^k=\tr\mcR_C^k \ ,
\ee
for any value of $k$ and $-\infty\leq t\leq+\infty$.

\subsection{Instanton numbers}

The instanton numbers, obtained by integrating the d-\emph{th} Chern character over the CY d-fold, 
\be
              N_d=\frac{1}{d!}\int_{CY_d}\tr\left(\frac{i\mcF}{2\pi}\right)^d \ ,
\ee
are of physical interest. We shall compute them here only for CY cone 2-folds and 3-folds, for these are the most relevant cases for heterotic compactifications. 

The instanton's curvature 2-form $\mcF(\mcA)$ is obtained using Eq.\eqref{eq:master_cond} and Eqs.\eqref{eq:F(t)}-\eqref{eq:beta(t)}. For the $n=1$ case we easily find that
\be\begin{split}
           \tr(\mcF\wedge\mcF) & = -16\left[2(\partial_t\varphi)^2+(\partial_te^{2\int\varphi})^2\right]\epsilon\wedge\bar{\epsilon}\wedge e\wedge\bar{e}\\
                                & = -16\partial_t(\partial_t+1)\varphi^2 dt\wedge\eta\wedge J \ .
\end{split}\ee
Then, the instanton number on a CY cone 2-fold can be computed in terms of the 2nd Chern character as
\be
           N_2=-\frac{1}{8\pi^2}\int_{CY_2}\tr\mcF^2=\frac{1}{2\pi^2}\int_{SE_3}\eta\wedge J \ .
\ee
Note that this computation is a mere check of well known results.

By contrast, the $n=2$ case that we shall treat now, is new. A straightforward but lengthy calculation leads to the following result
\be
           \tr\mcF^3=-i3\partial_t\varphi dt\wedge\eta\wedge\tr\mcR^2+i12f(t)dt\wedge\eta\wedge J^2 \ ,
\ee
where we introduced the $t$-dependent function
\be\begin{split}
           f(t)  = & 9\varphi^2e^{3\int\varphi}(\partial_t\varphi-1+\varphi+e^{3\int\varphi}) \\
& \hspace{18pt} -\partial_t\varphi\left(2(\partial_t\varphi)^2+6-6\varphi-2(1-\varphi)^2-2(1-\varphi)e^{3\int\varphi}+e^{6\int\varphi}\right)\ ,
\end{split}\ee
that vanishes both at the apex and infinity. We define the instanton number $N_3$ as the 3rd Chern character integrated over the CY$_3$ cone. We find it to be finite, being given by
\be
         N_3=\frac{-i}{48\pi^3}\int_{CY_3}\tr\mcF^3=-\frac{1}{8\pi}\int_{SE_5}\eta\wedge\tr\left(\frac{i\mcR}{2\pi}\right)^2+\frac{K}{4\pi^3}\int_{SE_5}\eta\wedge J^2 \ ,
\ee
where 
\be
            K=\int^{+\infty}_{-\infty} f(t) dt \ ,
\ee
is a finite number that does not depend on the choice of EK$_2$ base manifold. A numerical estimate gives $K=7/6$. Notice that given that the value of $K$ is the same for any CY cone 3-fold, the instanton number $N_3$ depends only on the geometric and topological properties of the base spaces, namely the volume of the SE$_5$ space and the volume and Euler number of the EK$_2$ base space.

\section{Examples}\label{sec:examples}

In this section we work out three concrete examples. We start by discussing the SU($2$) instanton over the CY cone 2-fold $\mathbb{C}^2/\mathbb{Z}_2$. We then move on to consider SU($3$) instantons over two CY cone 3-folds: the $\mathbb{C}^3/\mathbb{Z}_3$-orbifold and the conifold.

\subsection{The $\mathbb{C}^2/\mathbb{Z}_2$ orbifold}

One useful way of regarding the $\mathbb{C}^2/\mathbb{Z}_2$-orbifold is as being obtained by letting the single non-trivial compact 2-cycle on the Eguchi-Hanson space \cite{eguchi} collapse. Explicit SU($N$) HYM instantons over the latter have been constructed a long time ago, exploring the fact that the Eguchi-Hanson space is 4d hyperk\"ahler and the hermitian Yang-Mills instantons are self-dual. In particular, the SU(2) instanton that we now construct along the lines of the previous section has been known in the literature already for a while \cite{Bianchi:1996zj}.

The $\mathbb{C}^2/\mathbb{Z}_2$ orbifold is a cone over the Sasaki-Einstein $S^3/\mathbb{Z}_2$, that itself is a U(1) bundle over $\mathbb{CP}^1$. The Einstein-K\"ahler potential for the latter is
\be
                  \mcK=\half\ln(1+|w|^2) \ ,
\ee
and the \emph{einbein} reads
\be
                  e=\frac{1}{\sqrt{2}}\frac{dw}{1+|w|^2} \ .
\ee
Specializing the general solution of the previous section to the present case we obtain
\be
                  \mcA=-\half\frac{dz}{z}\sigma_3+\frac{1}{1+e^{t-t_0}}\left(\begin{array}{cc}
                      \half\epsilon& ~0 \\
                     -\sqrt{2}e & -\half\epsilon
                    \end{array} \right) \ .
\ee 
In a more convenient gauge, obtained via a $\phi$-dependent gauge transformation $U=e^{i\phi\sigma_3}$, the connection can also be written as
\be
                  \mcA'=U^{-1}\mcA U+U^{-1}\partial U=\frac{1}{1+e^{t-t_0}}\left(\begin{array}{cc}
                      \half\epsilon& ~0 \\
                     -\sqrt{2}e' & -\half\epsilon
                    \end{array} \right) \ .
\ee 
Here, we used a different but equivalent einbein $e'=e^{2i\phi}e$. Let us also write down the curvature 
\be\label{eq:curvature_n=2}
                  \mcF(\mcA)=\frac{e^{t-t_0}}{(1+e^{t-t_0})^2}\left(\begin{array}{cc}
                      \epsilon\wedge\bar{\epsilon}-2e\wedge\bar{e}&  -2\sqrt{2}\epsilon\wedge\bar{e} \\
                     -2\sqrt{2}e\wedge\bar{\epsilon} & -\epsilon\wedge\bar{\epsilon}+2e\wedge\bar{e}
                    \end{array} \right) \ .
\ee
($ \mcF(\mcA')$ can be obtained from \eqref{eq:curvature_n=2} replacing $e$ by $e'$.) We notice that $\mcF(\mcA)$ approaches $\mcR_C=0$ both at the apex of the cone and at infinity. This is not the case for $n\neq 1$. The instanton number can be explicitly shown to agree with the calculation of the previous section
\be
              N_2=\frac{1}{2\pi^2}\int_{S^3/{\mathbb Z}_2} \eta\wedge J=-\frac{1}{2} \ .
\ee
To obtain this result we used that on the $S^3/{\mathbb Z}_2$ orbifold, the periodicity of $\phi$ is halved, $\phi\sim\phi+\pi$.

It would be interesting to consider also SU($2$) instantons over the Eguchi-Hanson space, the blowup of the $\mathbb{C}^2/\mathbb{Z}_2$ orbifold. These are well-known, albeit not in a form that makes the complex structure evident.

\subsection{The $\mathbb{C}^3/\mathbb{Z}_3$ orbifold}

The next example is the $\mathbb{C}^3/\mathbb{Z}_3$ orbifold. This CY cone 3-fold has a $\mathbb{CP}^2$ EK base, the K\"ahler potential of the latter being
\be
                 \mcK=\frac{1}{2}\ln\left(1+|w^1|^2+|w^2|^2\right) \ .
\ee
We do not need to compute the spin-connection for this cone, for we know it to be flat outside the orbifold singularity, hence $\mcR_C=0$. (For more details on the $\mathbb{C}^n/\mathbb{Z}_n$-orbifolds and their resolutions we refer the interested reader to \cite{walter} and references therein.) It readily follows that 
\be
                 \tr\mcR^2=-12J^2 \ ,
\ee
and thus, the instanton number reads
\be
                 N_3=\frac{2K-3}{8\pi^3}\int_{S^5/Z_3}\eta\wedge J^2=\frac{2K-3}{48} \ .
\ee

\subsection{The conifold}

The final example of this section is an SU(3) HYM instanton over the conifold. We recall that the conifold is a cone over the Einstein-Sasaki manifold $T^{1,1}=(S^3\times S^3)/$U(1), that in turn is a U(1) bundle over EK$_2=\mathbb{CP}^1\times\mathbb{CP}^1$. The K\"ahler potential on the base can be taken to be
\be
            \mcK=\frac{1}{3}\ln(1+|w^1|^2)+\frac{1}{3}\ln(1+|w^2|^2) \ ,
\ee
and the resulting EK$_2$ metric is manifestly toric. It is useful to introduce the \emph{zweibein}
\be
            e_i=\frac{1}{\sqrt{3}}\frac{dw^i}{1+|w^i|^2} \ ,
\ee
in terms of which the (holomorphic part of the) spin-connection on $\mathbb{CP}^1\times\mathbb{CP}^1$ can be written as 
\be
        \mcC=-\sqrt{3}\left(\begin{array}{cc}
              \bw^1e_1 &\quad 0 \\
              0 & \bw^2e_2
              \end{array} \right) \ ,
\ee
leading to
\be
        \mcR=6\left(\begin{array}{cc}
              e_1\wedge\bar{e}_1 &\quad 0 \\
              0 & e_2\wedge\bar{e}_2
              \end{array} \right) \ .
\ee
It follows that $\mcR ic=i\tr(\mcR)=-6J$, as expected. Computing the conifold's spin-connection and curvature is also straightforward. We find
\be
        \mcA_C=\frac{1}{\sqrt{3}}\left(\begin{array}{ccc}
              \bw^ie_i &\quad 0 &\quad 0 \\
              -\sqrt{6}e_1  & -2\bw^1e_1+\bw^2e_2 & \quad 0 \\
              -\sqrt{6}e_2 &\quad 0 & \bw^1e_1-2\bw^2e_2
              \end{array} \right) \ ,
\ee
and
\be
        \mcR_C=2\left(\begin{array}{ccc}
              0 &\quad 0 &\quad 0 \\
              0  &\quad e_1\wedge\bar{e}_1-e_2\wedge\bar{e}_2 & -e_1\wedge\bar{e}_2 \\
              0 &-e_2\wedge\bar{e}_1  & -e_1\wedge\bar{e}_1+e_2\wedge\bar{e}_2
              \end{array} \right) \ .
\ee
One can easily check that $J_C^2\wedge\mcR_C=0$. The following identities are also simple to obtain:
\be
       \tr\mcR^2=0 \ ,
\ee
\be
       \tr\mcR_C^2=12J^2 \ .
\ee

In the present case, we \emph{can} consider a more general ansatz for $\Phi$ than in section \ref{sec:SU(n)instantons},
\be
             \Phi=\left(\begin{array}{ccc}
              \varphi_1(t)+\varphi_2(t) &\quad 0 &\quad 0 \\
              0  & -\varphi_1(t) & \quad 0 \\
              0 & 0  & -\varphi_2(t)
              \end{array} \right) \ .
\ee
As before, we take $\alpha=\partial\mcK$ and $\Gamma=\mcC+(\partial\mcK)\mathbb{1}$, while 
\be
             E_i=-\sqrt{2}e^{\int^t_{t_i}(\varphi_i+\varphi_1+\varphi_2)dt}e_i \ .
\ee
With these choices, $\Phi$ and $\mcB$ satisfy the holomorphy conditions \eqref{eq:Fterm_3a} and \eqref{eq:Fterm_3b}. The DUY equation then leads to two coupled differential equations for $\varphi_{1,2}(t)$:
\be\label{eq:DUY_conifold}
                 2\partial_t\varphi_i+4\varphi_i+1-e^{2\int^t_{t_i}(\varphi_i+\varphi_1+\varphi_2)dt}=0 \ .
\ee
Imposing the condition that at the IR (i.e. $t=-\infty$) one has $\partial_t\varphi_i=0$, it follows from Eq.\eqref{eq:DUY_conifold} both that $t_i=-\infty$ and $\varphi_i^{IR}=0$. Introducing the variables 
\be
             X_i=2\int^t_{-\infty}(\varphi_i+\varphi_1+\varphi_2)dt \ ,
\ee
we can now recast the flow equations \eqref{eq:DUY_conifold} as 
\be
            \partial_t^2X_i+2\partial_tX_i=-G_{ij}\frac{d}{dX_j}(X_1+X_2-e^{X_1}-e^{X_2}) \ ,
\ee
with metric
\be
            G_{ij}=\left(\begin{array}{cc}
               2 & 1 \\
              1 & 2
              \end{array} \right) \ .
\ee
The above equation describes the damped motion of a particle in a 2D space with  metric $G_{ij}$, in the potential $V=X_1+X_2-e^{X_1}-e^{X_2}$. By definition we have $X_i^{IR}=0$, and inspection of the above equations for $X_i\simeq 0$ shows that there are only four branches of solutions satisfying these boundary conditions. In fact, we find that for $t\ll 0$ either $X_1=X_2$ or $X_1=-X_2$. Since only solutions with $X_i\leq 0$ lead to finite instanton numbers, we see that the only interesting branch of solutions is the one with $X_1=X_2\leq 0$ at $t\ll 0$. In fact, the symmetry under interchange of $X_1$ and $X_2$ implies then that $X_1=X_2$ for all values of $t$, and we are back to the solutions discussed in section \ref{sec:SU(n)instantons}, with $\varphi_1=\varphi_2=\varphi/2$. Given that $\tr\mcR^2=0$, the instanton number of our SU($3$) instanton over the conifold is
\be
          N_3=\frac{K}{4\pi^3}\int_{T^{1,1}}\eta\wedge J^2=\frac{2K}{27} \ .
\ee
where we used that $\phi\sim\phi+4\pi/3$.

\section{Future directions}

We close with a short discussion of open problems and future work. In this paper, we considered a natural ansatz that reformulates the hermitean Yang-Mills equations on Calabi-Yau cones in terms of a suitable Higgs-Yang-Mills flow on Einstein-K\"ahler manifolds. This new perspective helped us finding - for any CY$_d$ cone - a family of SU($d$) instantons that depend explicitly on a single parameter, the size modulus $t_0$. One expects however more general multi-parameter solutions to exist, that should be connected to our instantons by smooth deformations. In fact, it is known that the instanton moduli spaces should be K\"ahler, hence at least two-dimensional. Typically, at least some of the extra moduli that we are missing here should be related to global gauge transformations. It is an interesting open problem to find the number of these moduli and understand the geometry of the instanton moduli space $\mcM$. A na\"ive computation shows that the metric (and the K\"ahler potential) of $\mcM$ should be proportional to $e^{nt_0}$. This leads us to conjecture that $\mcM$ should itself be a K\"ahler cone, with $\rho_0=e^{nt_0/2}$ playing the r\^ole of the radial coordinate. We will report on this issue in a forthcoming publication \cite{paccetti2009}.

As we have already mentioned, there is a large family \cite{Martelli:2007pv} of CY cones that can be (at least partially) smoothed out. It is an interesting problem to find out if the construction we presented herein can be easily modified to take into account the deformation of the geometry at the tip of the cones (i.e. at the IR). In particular, one might wonder if there is a family of instantons that converges to the one we found in the present work as we take the singular limit on the CY geometry. Explicit instantons on these smooth non-compact CY's should be of interest not only for \emph{computable} toy models for heterotic supergravity compactifications, but also in the context of the gauge/gravity duality. Several properties of the 4d field theory duals should be intimately related to the features of the Higgs-Yangs-Mills flow we described in this paper.

\section*{Acknowledgements}
The author is grateful to S. Groot Nibbelink, C. A. R. Herdeiro and M. G. Schmidt for useful discussions. It is a pleasure to acknowledge the kind hospitality of the Institut f\"ur Theoretische Physik, Heidelberg, Germany, during a visit in the early stage of this work. The work of F.P.C. is supported by \emph{Funda\c c\~ao para a Ci\^encia e a Tecnologia} through the grant SFRH/BPD/20667/2004.

\appendix

\section{More general instantons?}

In the following we shall consider the possibility of finding 
SU($n+1$) instantons on CY$_{n+1}$ cones with the ansatz of section \ref{sec:SU(n)instantons} other than the ones already discussed therein. For convenience, we write
\be
                 \mcB=\left(\begin{array}{cc}
              \alpha & 0 \\
              E & \Gamma_{n\times n}
              \end{array} \right) \ .
\ee
with
\be
                 E_i=-c_0\sqrt{2}\exp{\left(\frac{n+1}{n}\int^{~t}_{-\infty}\varphi dt\right)}~E_i^0 \ ,
\ee
and recall that $E_i^0$, $\alpha$ and $\Gamma_{ij}$ are $t$-independent. Once the holomorphy condition is satisfied at a given $t$, it is automatically satisfied also at any other $t$. Actually, one can multiply $E_i$ by any ($t$-dependent) number to get a new holomorphic connection. 

Assuming the above connection to be holomorphic, we now want to learn which constraints follow from the DUY equation. It is not difficult to find the following two conditions:
\be\label{eq:cond_1}
                \left(\bar{\partial}E^0_i+(\bar{\Gamma}_{ij}-\delta_{ij}\bar{\alpha})\wedge E^0_j\right)\wedge J^{n-1}=0 \ ,
\ee
and
\be\label{eq:cond_2}
                E^0_i\wedge\bar{E}^0_j\wedge J^{n-1}\propto \delta_{ij}J^n \ .
\ee
Eq.\eqref{eq:cond_2} implies that the n-vector 1-form $E_i^0$ is related to the n-bein $e_i$ by a U($n$) transformation. If this is the case, then by suitably rescaling $c_0$ and changing the choice of orthonormal n-bein we can safely set $E^0_i=e_i$. 

Let us introduce the connection $\mcB'$ defined as
\be
                 \mcB'=\left(\begin{array}{cc}
              \alpha & 0 \\
              -\sqrt{2}e & \Gamma_{n\times n}
              \end{array} \right) \ .
\ee
According to the previous discussion, if $\mcB$ is holomorphic as we assume, then $\mcB'$ is holomorphic too. On the other hand, we see that the DUY implies also that 
\be\label{eq:cond_3}
               \mcF_{1,1}(B')\wedge J^{n-1}=id_0\Sigma J^n \ ,
\ee
where 
\be
               d_0=\frac{1}{2(n+1)}\tr(\mcK^{i\bj}\mcF_{i\bj}(B')\Sigma) \ .
\ee
Finally, in case $\mcB'$ is holomorphic and it also satisfies \eqref{eq:cond_3}, we can find an SU($n+1$) instanton by solving the flow equation
\be
                 \partial_t\varphi+n\varphi+\frac{n}{2}(1-d_0)-\frac{n}{2}c_0^2\exp{\left(2\frac{n+1}{n}\int^{~t}_{-\infty}\varphi~dt\right)}=0 \ .
\ee
Requiring that the solutions of this equation correspond to bounded instantons leads to the condition that $d_0<1$. In this case, this flow equation is quite similar to the one we studied in section \ref{sec:SU(n)instantons}. We can use the same type of reasoning we used there to argue that there are bounded solutions only if $c_0^2= 1-d_0$. These solutions have $\partial_t\varphi=0$ both at the IR and the UV and
\be
                \varphi^{IR}=0 \ , \quad \varphi^{UV}=-\frac{c_0^2}{2} \ .
\ee
This implies in particular that 
\be
                \mcK^{i\bj}\mcF_{i\bj}(B^{IR})=0 \ ,
\ee
that is, the connection $\mcB$ approaches a HYM connection over the EK$_n$ base at the IR. The remaining parameter $c_0^2$ is determined by $\mcB^{UV}$ as
\be
              c_0^2=-\frac{1}{2(n+1)}\tr(\mcK^{i\bj}\mcF_{i\bj}(B^{UV})\Sigma) \ .
\ee

The upshot of this discussion is that any bounded instanton constructed using the ansatz of section \ref{sec:SU(n)instantons} is described by a Higgs-Yang-Mills flow that in the IR approaches a HYM instanton over the EK$_n$ base (and therefore also $\Phi^{IR}=0$). In this paper we were able to explicitly construct instantons that approach the spin-connection at the apex of the cone. However, the analysis of the present appendix suggests that other solutions (obeying this ansatz) might exist that approach other HYM instantons over the EK$_n$ at the IR. Let us recall the conditions imposed on these instantons. Writing $\Gamma=\mcC'+\mathbb{1}\partial\alpha$ these can be summarised as
\be
                   \mcF_{2,0}(\mcC')=0 \ , \quad \mcK^{i\bj}\mcF_{i\bj}(C')=2(n+1)c_0^2\mathbb{1} \ ,
\ee
and
\be\label{eq:cond_fin}
                   \partial e+\mcC'\wedge e=0 \ , \quad (\bar{\partial} e+\bar{\mcC}'\wedge e)\wedge J^{n-1}=0 \ .
\ee
That is, our problem boils down to a search for an U($n$) HYM instanton on the EK$_n$ base, such that its connection $C'$ satisfies \eqref{eq:cond_fin}, and differs from the spin-connection (on the EK$_n$ space). The constant $c_0^2$ is (up to a constant) the first Chern number of the U($n$) bundle with connection $C'$.

\end{document}